# Multiphase structure of finite-temperature phase diagram of the Blume-Capel model. Wang-Landau sampling method.

**G. Pawłowski**

UAM Institute of Physics, Umultowska 85, 61-614 Poznań, Poland

We investigate the density of states (DOS) in an antiferromagnetic spin-system on a square lattice described by the Blume-Capel (BC) model. We use a new and very efficient simulation method, proposed by Wang and Landau, in which we estimate very precisely DOS by sampling in the space of energy. Then we calculate the thermodynamical averages like internal energy, free energy, specific heat and entropy.

The BC model exhibits multicritical behaviour such as first- or second-order transitions and tricritical points. It is known that the ground state of the model can exhibit two kinds of staggered antiferromagnetic phases: AF$_1$ (two interpenetrating lattices with S = -1 and S = 1) and AF$_2$ (S = -1 and S = 0 for *H < 0*; S = 1 and S = 0 for *H > 0*). We analyze the coexistence of such phases at finite temperatures and determine border lines between them. To understand the microscopic nature of such boundaries we present also some results obtained with the standard Monte Carlo method.

**1 Introduction** The Blume-Capel (BC) model was introduced [1, 2] to describe magnetic materials with single-ion anisotropy. The Hamiltonian of the model is given by:

$$\hat{H}_{BC} = -J \sum_{<i,j>} S_i S_j + D \sum_i S_i^2 \qquad (1)$$

where $S_i$ is the spin-1 Ising operator ($S_i = \pm 1, 0$), $J$ is the spin-spin interaction between nearest-neighbors and $D$ is the single-ion anisotropy. For $J < 0$ (antiferromagnetic case) this model exhibits multicritical behaviour such as first- or second-order transitions and tricritical points. In the ground state one can observe 1$^{st}$ order transition from (-1,1) antiferromagnetic phase to (0,0) paramagnetic state for $D/|J| = 2.0$. The results of Monte Carlo simulations [3-6] and analysis by very accurate numerical methods [7] indicate the tricritical point presence for $D/|J| = 1.965(5)$ and $k_B T/|J| = 0.609(4)$.

It is very interesting to analyze properties of such a model in the presence of an external magnetic field *H*:

$$\hat{H} = \hat{H}_{BC} - H \sum_i S_i . \qquad (2)$$

First of all, in such a case in the ground state we observe two kinds of ordered phases: for small values of *H* there exists the standard Ising S=1 (1,-1) antiferromagnetic phase, which we call in this paper AF$_1$, but for larger values of the magnetic field we can also observe the second antiferromagnetic phase (0, ±1) (respectively for H>0 or H<0, which we call here AF$_2$). In the ground state there is the discontinuous order-order transition (described by the equation $H = 4|J| - D$ for $0 \leq D/|J| \leq 2$). It has been suggested that due to the absence of next-nearest neighbor interactions, the boundary separating the ordered phases is present in the ground state only [3, 4].

We have made two independent analyses of antiferromagnetic states in finite temperatures and we have found the decomposition of the order-order critical point in the ground state to two boundaries. As the first step we perform standard Monte Carlo simulations and analyze the 'infinite clusters' of $AF_1$ and $AF_2$ (we mark out the percolation boundaries for *contour clusters*, in the sense of definitions in Ref. [8]). The second, completely different method is based on the idea of Wang and Landau (WL) [9]. We simulate very accurate density of states (DOS) and then we solve thermodynamical properties of the model. The obtained plots of specific heat and internal energy as a function of temperature confirm the existence of $AF_1$ and $AF_2$ at low nonzero temperatures.

**2 Infinite clusters** Here we present results of standard Monte Carlo simulations, using the Metropolis algorithm for a square lattice L=64 with the periodic boundary conditions. The analysis for larger systems 80x80, 120x120 has been also performed and the results have been qualitatively the same. In Fig. 1 we present the phase diagram of the BC model for $D/|J|=1$ as a function of $H/|J|$ and $k_B T/|J|$. The solid line indicates 2$^{nd}$ order antiferromagnetic-paramagnetic phase transitions, which is in good agreement with previous results [3, 4]. This boundary is obtained in the standard way by measurement of fluctuations of internal energy (specific heat) and staggered magnetization (staggered susceptibility), peaks in the above quantities are interpreted as phase transitions.

The crucial question of our analysis is 'how to obtain detailed information about the type of antiferromagnetism in the ordered state?' We have analysed existing *ordered contour clusters* in each accepted MC sample. Especially we searched for the 'infinite' (spanning) contour cluster (note: here we do not construct a 'physical' geometric clusters as in Ref. [6] and [10]). Using the percolation cluster algorithm [11] we obtain information about percolation transition in the system. From the detailed analysis of spanning cluster probability $P_S^{phase}$, which we define as the ratio of the number of states with spanning cluster for *phase* and the total number of states generated and accepted for one Monte Carlo point (simulation for fixed $D/|J|$, $H/|J|$ and temperature), we determine the temperatures at which the percolation transition (PT) takes place $T_{PT}^{phase}(P_S^{phase})$ (where $0 < P_S^{phase} < 1$).

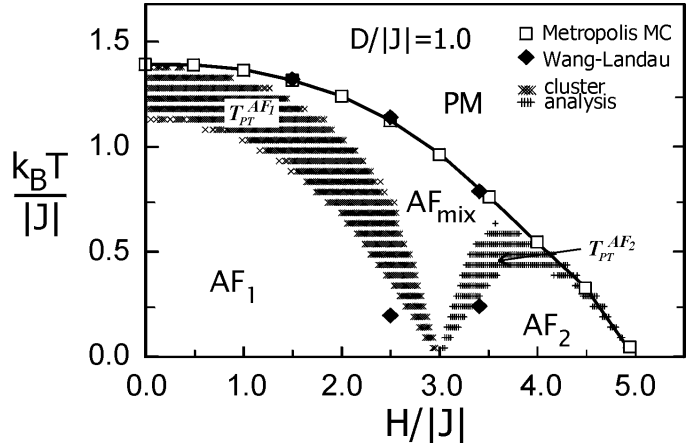

**Fig. 1** Phase diagram of the BC model for a square lattice for $D/|J|=1$ as a function of external magnetic field $H/|J|$. The solid line, which indicates 2$^{nd}$ order antiferromagnetic-paramagnetic transition, was obtained using the standard Metropolis MC (□; L=64) and WL (◆; L=20, cf. Fig. 2) simulations. Horizontal lines indicate percolation transition for *contour clusters* of $AF_1$ and $AF_2$ (for the area '$AF_1$' and '$AF_2$' the condition $P_S^{phase} = 1$ is satisfied, while for '$AF_{mix}$': $P_S^{phase} = 0$). Denotations: $AF_{mix}$ – antiferromagnetic mixed phase, which is contained of ordered structures of $AF_1$ and $AF_2$, PM – paramagnetic phase.

The horizontal lines in Fig. 1 indicate a percolation transition area for a *contour cluster* of ordered phases. We see that also $AF_1$ and $AF_2$ states exist in finite temperatures. The 1$^{st}$ order $AF_1$-$AF_2$ transition at finite temperatures disappears, and we observe two boundaries $AF_1$-$AF_{MIX}$ and $AF_{MIX}$-$AF_2$. Using standard Monte Carlo method we cannot specify definitively the character of the boundaries observed.

**3 The Wang-Landau method** The second way to analyse thermodynamic properties of the system is the simulation method, proposed by Wang and Landau, in which we estimate very precisely the density of states by sampling in the space of energy. This method is based on the idea of flat histogram, which we build during a random walk in the energy space with a probability proportional to the reciprocal of the density of states $1/g(E)$ [9]. After many cycles, in which the accuracy factor $f$ is modified, we obtain a reliable density of states in the system analysed. In this paper we use the original algorithm proposed by Wang and Landau and implement also a modification proposed by Zhou and Bhatt [12]. The results presented have been obtained with the final factor $\ln(f) \approx 10^{-9}$. After building a flat histogram for a given accuracy, we calculate the thermodynamical averages like internal energy, free energy, specific heat and entropy.

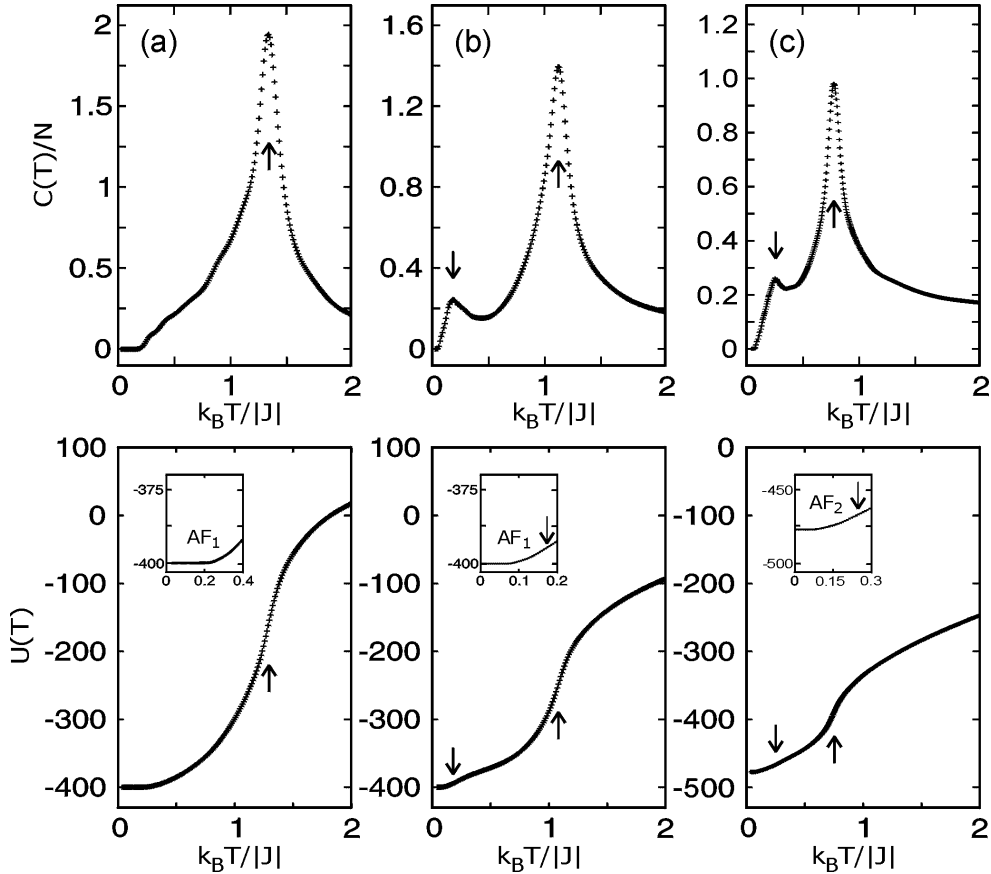

**Fig. 2** Specific heat and internal energy for the BC model (L=20, square lattice) obtained using the Wang-Landau algorithm for $D/|J|=1$ as a function of temperature. Columns present results for (a) $H/|J|=1.5$; (b) 2.5 and (c) 3.4, respectively. Upward arrows indicate the phase transitions, while downward arrows the order-order boundaries. In boxes the internal energy near the ground state is presented (existence of overall ordered domain corresponds to a constant value of $U(k_BT)=U(0)$).

Fig. 2 presents the results of the calculations (based on the accurate DOS data) of the specific heat and internal energy as a function of temperature for the BC model for $D/|J|=1.0$ and three selected values of $H/|J|$. In the column (a) and (b) in Fig. 2 the characteristics of the system for $H/|J|$ = 1.5 and 2.5 are presented. On the specific heat characteristics $C(T)/N$ we observe well-marked peaks of second order antiferromagnetic-paramagnetic transition. On the internal energy plot we observe a constant value of energy (U = -400 for L=20) near the ground state, for $H/|J|$ = 1.5 up to $k_BT/|J|$ = 0.25 and for $H/|J|$ = 2.5 up to $k_BT/|J|$ = 0.09 (cf. boxes in Fig 2a,b); it is a clear manifestation of the AF$_1$ ordered phase existence in low temperatures. With increasing temperature the internal energy grows because the overall size of ordered cluster decreases, although the domain of AF$_1$ is still infinite (spanning) up to the temperatures given by $T_{PT}^{phase}(P_S^{phase})$. An analogous situation is observed for the AF$_2$ phase; $H/|J|$ = 3.4 (cf. Fig. 2c).

For $H/|J|$ = 2.5 and 3.4 we can observe also a local maximum of specific heat at low temperatures. This anomalies are connected with transformation between the orders AF$_1$-AF$_{MIX}$ and AF$_2$-AF$_{MIX}$, but we cannot explicitly interpreted them as an order-order transitions (the transition effect on specific heat and internal energy is very weak, Figs. 1, 2bc). There are two questions: what is the role of the finite size effects on observed transition between AF$_1$-AF$_{MIX}$ and AF$_2$-AF$_{MIX}$ and why we do not observe such a transition in the whole range of external magnetic field (cf. Fig. 2a)?. An open question is also the coincidence between percolation transition (percolation threshold) and the specific heat anomalies in the thermodynamic limit.

**4 Summary**  We present new results concerning the multiphase structure of finite-temperature phase diagram of the Blume-Capel model obtained using standard Monte Carlo simulations as well Wang-Landau flat-histogram method. We show the existence of three antiferromagnetic ordered phases (AF$_1$, AF$_2$ and AF$_{MIX}$) in finite temperatures and the order-order boundaries between them in low temperatures. It will be very interesting to analyse such a problem in the range of model parameters covering the 1$^{st}$ order transitions to paramagnetic phase using complementary cluster approaches [13].

**Acknowledgements**  Author would like to thank S.Robaszkiewicz for many useful discussions. This work was supported in part by the Polish State Committee for Scientific Research (KBN), Grant No: 1 P03B 084 26, and by the Foundation for Polish Science (professors subsidy, R.Micnas).